\documentclass[aps,prl,twocolumn,showpacs]{revtex4}
\usepackage{amsmath,amsfonts,amssymb,graphics,graphicx,epsfig,color,bbm,subfigure}
\usepackage{fancyhdr}

\newcommand{\R}{\mathbbm{R}}

\newcommand{\tr}{{\rm Tr}\,}
\renewcommand{\det}{{\rm Det}\,}
\newcommand{\gr}[1]{\boldsymbol{#1}}
\newcommand{\be}{\begin{equation}}
\newcommand{\ee}{\end{equation}}
\newcommand{\bea}{\begin{eqnarray}}
\newcommand{\eea}{\end{eqnarray}}

\newcommand{\eq}[1]{Eq.~(\ref{#1})}

\begin{document}

\title{Multimode uncertainty relations and separability of continuous variable states}
\author{Alessio Serafini}
\affiliation{Department of Physics \& Astronomy, University College London, 
Gower Street, London WC1E 6BT, United Kingdom}
\affiliation{Dipartimento di Fisica ``E.R. Caianiello'', Universit\`a di Salerno, Via S.~Allende, 
84081, Baronissi, Italy}
\begin{abstract}
A multimode uncertainty relation (generalising the Robertson-Schr\"odinger relation) is derived as a 
necessary constraint on the second moments of $n$ pairs of canonical operators. 
In turn, necessary conditions for the separability of multimode continuous variable states
under $(m+n)$-mode bipartitions are derived from the uncertainty relation. 
These conditions are proven to be necessary and sufficient for $(1+n)$-mode Gaussian states 
and for $(m+n)$-mode bisymmetric Gaussian states. 
\end{abstract}

\maketitle

Quantum mechanical uncertainty principles stem directly 
from the noncommutativity of quantum observables and
from the probabilistic interpretation of the wavefunction.  
In a close sense, uncertainty relations \cite{heisenberg27,robertson29,dodonov80}
are {\em the} `operational' expressions 
of such fundamental axiomatic features.
Besides this outstanding fundamental interest, uncertainty relations 
have recently acquired a technical interest of major relevance for
quantum information science \cite{guehneprl04}. 
In fact, for any state $\varrho$ of a bipartite quantum system, 
the positivity of the partially transposed density matrix $\tilde{\varrho}$ 
(obtained from $\varrho$ by transposing the Hilbert space of only one of the two 
subsystems) is a necessary condition for the state to be separable \cite{peres96,horodecki96}. 
In other words, the violation of the positivity of $\tilde{\varrho}$ is a proof of the presence 
of quantum entanglement in the state $\varrho$, 
which is thus suitable for various quantum informational aims.
Now, because uncertainty relations for quantum observables 
derive from the positivity of the generic density matrix $\varrho$, 
the partial transposition of any uncertainty relation ({\em i.e.~}the relation formally 
derived from the uncertainty relation by considering the partially transposed state $\tilde{\varrho}$
instead of $\varrho$) provides a sufficient condition for the state $\varrho$ to be entangled.
Often, as is the case for the continuous variable systems to which this paper is devoted, 
such conditions come in a simple form in terms of observable quantities, and are therefore 
of great experimental relevance \cite{vanloock03}.
Furthermore, these conditions turn out to be also necessary for 
entanglement in several instances, namely whenever 
the positivity of the partial transpose is also
sufficient for separability \cite{horodecki96,simon00,werewolf,serafozzi05}.
Notably, the relationship between uncertainty relations and entanglement 
is exhibited in discrete variable systems as well (see Refs.~\cite{guehneprl04,entropic} 
and, for `spin-squeezed' states, \cite{sanders05}).

In this paper, canonical systems of many modes (like discrete bosonic fields in second quantization or 
motional degrees of freedom of material particles in first quantization) are considered. 
Quantities invariant under linear canonical ({\em i.e.~}symplectic) 
operations on the field modes will be constructed as functions of the second moments of the field operators.
In terms of such invariant quantities, a general uncertainty relation for the second moments of 
any $n$-mode system will be derived. 
The partial transposition of such a relation will lead to useful
entanglement conditions for any bipartition of the modes.  
Crucially, these multimode conditions are both
easily checkable and 
straightforwardly related to experimental data.

{\em Uncertainty relations for canonical systems. -- }
Let us consider a continuous variable (CV) quantum mechanical system 
described by $n$ pairs of canonically conjugated 
operators $\{\hat x_j,\hat p_j\}$ with continuous spectra.
Grouping the canonical operators together in the 
vector $\hat R=(x_1,p_1,\ldots,x_n,p_n)^{\sf T}$ allows to 
compactly express the canonical commutation relations (CCR) as
$
[\hat R_j,\hat R_k] = 2i\,\Omega_{jk} 
$
\cite{ccrnote}, 
where the symplectic form $\Omega$ is defined as 
$\Omega \equiv \oplus_{1}^{n} \omega$ with 
$\omega\equiv i s_{y}$ ($s_{y}$ 
standing for the $y$ Pauli matrix).
Dynamical evolutions of the system must preserve the CCR.
In particular, transformations acting {\em linearly} 
on the vector of operators $\hat R$ (in Heisenberg picture)
must preserve the symplectic form $\Omega$ under congruence. 
Such 
transformations form the {\em real symplectic group} $Sp_{2n,\R}$: 
$S\in SL({2n,\R}) \, :\;
S\in Sp_{2n,\R} \Leftrightarrow S^{\sf T} \Omega S = \Omega$. 
These transformations can be surjectively mapped onto unitary operations
generated by second order polynomials in the canonical operators 
({\em metaplectic} representation).
 
Any state of an $n$-mode CV system is described by a positive, trace-class 
operator $\varrho$. Let us define the $2n\times 2n$ matrix of second moments, 
or ``covariance matrix'' (CM), 
${\gr\sigma}$ (with entries $\sigma_{i,j}$) of the state $\varrho$ as
${\sigma}_{i,j} \equiv \tr{[\{\hat R_i , \hat R_j\} \varrho]}/2
-\tr{[\hat R_i \varrho]}\tr{[\hat R_j \varrho]}$.
The full uncertainty relation for such a system reads \cite{simon8794}
\be
{\gr\sigma} + i\Omega \ge 0 \; . \label{obsheis} 
\ee
This inequality -- {\em which derives solely from the CCR and 
from $\varrho\ge0$} -- 
is the only condition a symmetric $2n\times 2n$ matrix has to satisfy 
to qualify as the {\em bona fide} CM of a physical state. 
Because of the skew-symmetry of $\Omega$, Eq.~(\ref{obsheis}) ensures 
the definite positivity of ${\gr\sigma}$: ${\gr\sigma} > 0$. 
For future convenience, let us define the 
$2\times2$ submatrices ${\gr\gamma}_{ij}$ [with entries $(\gamma_{ij})_{h,k}$] 
of the CM $\gr\sigma$ as $(\gamma_{ij})_{h,k}\equiv\sigma_{(i+h-1),(j+k-1)}$
(each of them describing one mode or the correlations between one pair of modes).

For $n=1$, the uncertainty principle (\ref{obsheis}) reduces 
to $\det{{\gr\sigma}}\ge 1$ 
({\em i.e.~}to the ``Robertson-Schr\"odinger'' uncertainty relation
\cite{robertson29}). 
This relation along with the condition $\gr\sigma > 0$ are {\em equivalent} 
to the uncertainty relation (\ref{obsheis}) for single-mode systems. 
For a two-mode system, one has
\be
\det{{\gr\sigma}} + 1 \ge \Delta_{1} \; , \label{sera}
\ee
where $\Delta_{1}\equiv\sum_{i,j=1}^{2}\det{{\gr\gamma}_{ij}}$. 
Note that the quantities $\det{{\gr\sigma}}$ and $\Delta_{1}$, 
entering in the previous Inequalities, are
invariant under symplectic transformations \cite{serafozzi04}.
The invariant nature of the uncertainty principle implies that, indeed, 
the expression of the uncertainty relation for a general $n$-mode CM ${\gr\sigma}$
must be possible in terms of symplectic invariants constructed from 
the entries of ${\gr\sigma}$. 

{\em Construction of the symplectic invariants. -- } 
Let us recall that, because of the definite positivity of $\gr\sigma$,
one can apply a seminal result by Williamson \cite{williamson} to the quadratic form ${\gr\sigma}$ 
to infer the following basic result:  
for any CM ${\gr\sigma}$ there exists a (non-unique) symplectic transformation $S\in Sp_{2n,\R}$
such that 
$
S^{\sf T} {\gr\sigma} S = {\gr\nu} \; , 
$
where 
$
{\gr \nu} = \oplus_{j=1}^{n} \,{\rm diag}\,{(\nu_j,\nu_j)} \; . 
$
The quantities $\{\nu_j\}$ are referred to as {\em symplectic eigenvalues}, 
whereas the matrix ${\gr\nu}$ is the ``normal form'' of the CM ${\gr\sigma}$.   
The uncertainty principle (\ref{obsheis}) can be equivalently recast in terms 
of the $\{\nu_j\}$ as 
\be
\nu_j\ge 1 \quad {\rm for} \;\; j=1,\ldots,n \, . \label{eigheis}
\ee

The symplectic eigenvalues are clearly symplectic invariants but, for an $n$-mode
system, their analytical expression in terms of the second moments turns out to be 
rather cumbersome (when possible at all).
Indeed, the symplectic eigenvalues can be computed by diagonalising 
the matrix $\Omega{\gr\sigma}$, whose eigenvalues turn out to be $\{\mp i\nu_j\}$ 
for $j=1,\ldots,n$. The latter statement is easily proved by checking it on the normal form 
${\gr \nu}$ and by considering that 
$
\Omega{\gr\nu} = \Omega S^{\sf T}{\gr\sigma} S = S^{-1}\Omega{\gr\sigma} S 
$ 
for some $S\in Sp_{2n,\R}$. 
Now, a natural choice of symplectic invariants, dictated by the previous equation, is 
given by the principal minors of the matrix $\Omega{\gr\sigma}$, manifestly 
invariant under symplectic transformations acting by congruence on ${\gr\sigma}$.
Let $M_{k}(\alpha)$ be the principal minor of order $k$ of the matrix $\alpha$ \cite{minornote}, 
then the symplectic invariants of an $n$-mode state $\{\Delta^n_j\}$ 
for $j=1,\ldots,n$ are defined as
\be
\Delta^n_j \equiv M_{2j}(\Omega{\gr\sigma}) \; . \label{definva}
\ee
The principal minors of odd order vanish because of the alternate sign in the 
spectrum of $\Omega{\gr\sigma}$, thus leaving us with $n$ independent 
symplectic invariants $\{\Delta^n_j\}$
(as many as the symplectic eigenvalues).
The quantities $\{\Delta^{n}_j\}$ are also known as ``quantum universal invariants'' \cite{dodonov00}.
The expression of the invariants $\{\Delta^n_j\}$ in terms of 
the symplectic eigenvalues $\{\nu_j\}$ can be retrieved by 
considering the normal form ${\gr\nu}$ and reads
\be
\Delta^n_j = 
\sum_{{\cal S}^{n}_{j}} \prod_{k\in{\cal S}^{n}_{j}} \nu_{k}^{2} \; , \label{inva}
\ee
where the sum runs over all the possible $j$-subsets ${\cal S}^{n}_{j}$ 
of the first $n$ natural integers ({\em i.e.~}over all the possible combinations of 
$j$ integers smaller or equal than $n$). 
Clearly, one has $\Delta^n_n=\det{{\gr\sigma}}$ while, for two-mode states, 
the invariant $\Delta^2_1$ coincides with the quantity $\Delta_1$ appearing in the uncertainty 
relation (\ref{sera}). 

{\em Symplectic uncertainty relations. -- } 
Let us consider an $n$-mode CV system and define the quantity ${\Sigma}_n$ as
\be
{\Sigma}_n = \sum_{j=0}^{n} (-1)^{n+j} \Delta^n_j \; , \label{sig}
\ee
where we assume $\Delta^n_0 \equiv 1$. 
Now, Eqs.~(\ref{inva}) and (\ref{sig}) imply 
\be
{\Sigma}_n = \prod_{j=1}^{n} (\nu_j^2-1) \; . \label{veritas}
\ee
Thus, Inequality (\ref{eigheis}) leads to the following statement:
 
\noindent {\em Symplectic uncertainty relation.} {Let ${\gr\sigma}$ be the covariance matrix 
of an $n$-mode continuous variable state.
The symplectic invariant ${\Sigma}_n$, determined according to Eqs.~(\ref{definva}) and (\ref{sig}), 
fulfills the inequality 
\be
{\Sigma}_n \ge 0 \; .  \label{symheis}
\ee} \smallskip

Inequality (\ref{symheis}), reducing to the well known relations $\det{\gr\sigma}\ge1$ and
(\ref{sera}) for, respectively, $n=1$ and $n=2$, 
provides a general 
way of expressing a necessary uncertainty relation constraining the symplectic invariants. 
\eq{veritas} shows that, actually, Inequality (\ref{symheis}) is 
only necessary and not sufficient for the full uncertainty relation (\ref{eigheis})
to be satisfied, as it is not able to detect unphysical CMs for which an even number of 
symplectic eigenvalues violates Inequality (\ref{eigheis}). 
This impossibility is not due to any 
fundamental lack of information in the symplectic invariants, since their knowledge 
allows to determine 
the symplectic eigenvalues as the $n$ solutions for $\nu$ 
of the following system in the unknown $\{\Delta_j^{n-1},\nu\}$ (for $j=1,\ldots,n-1$):
$\Delta^n_j = \nu^2 \Delta_{j-1}^{n-1}+\Delta_j^{n-1}$ for 
$j=1,\ldots,n$ ($\Delta_{n}^{n-1}\equiv0$).
For $n=2$, 
the additional proviso 
$\det{\gr\sigma}\ge1$ is enough to rule out the undetectable case so that 
the uncertainty principle (\ref{obsheis}) is {\em equivalent},
for two-mode states, to the set of conditions
$\Delta^2_2-\Delta^2_1+\Delta^2_0 \ge 0 \, , \;
\Delta^2_2\ge 1 \, , \;
{\gr\sigma}>0$.  

{\em Symplectic separability criteria. -- }
The positivity of the partially transposed state (``PPT criterion'') 
is a necessary condition for the separability of any bipartite quantum state.
Conversely, the violation of such positivity is a sufficient condition 
for a quantum state to be entangled.
Moreover, as far as the CV systems here addressed are concerned, 
the PPT criterion turns out to be sufficient as well for the separability of $(1+n)$-mode
Gaussian states ({\em i.e.~}states with Gaussian Wigner function)
and of bisymmetric $(m+n)$-mode Gaussian states
(here and in what follows, we refer to a bipartite `$(m+n)$-mode' CV state as to a state
separated into a subsystem $A$ of $m$ modes and a  
subsystem $B$ of $n$ modes).
These facts come in especially handy for CV systems, as the action of partial transposition 
on covariance matrices is easily described. Let $\varrho$ be a $(m+n)$-mode bipartite CV state 
with $2(m+n)$-dimensional CM ${\gr\sigma}$. 
Then the CM $\tilde{{\gr\sigma}}$ of 
the partially transposed state $\tilde{\varrho}$ with respect to, say, subsystem $A$, 
is obtained by switching the signs of the $m$ momenta $\{p_{j}\}$ belonging to subsystem $A$:
$
\tilde{{\gr\sigma}} = T {\gr\sigma} T $ with 
$T \equiv \oplus_{1}^{m} s_z \oplus {\mathbbm 1}_{2n}$, 
where ${\mathbbm 1}_{2n}$ and $s_z$ stand for the $2n\times2n$ identity matrix and 
for the $z$ Pauli matrix.
Now, in analogy with Inequality (\ref{obsheis}) 
derived from the positivity of the density matrix $\varrho$, 
a (generally) sufficient condition for separability derived by the PPT 
criterion is given by 
$\tilde{{\gr\sigma}}+i\Omega \ge 0$ \cite{simon00} 
or, in terms of the symplectic eigenvalues $\{\tilde{\nu}_j\}$ 
of the partially transposed CM $\tilde{{\gr\sigma}}$
(whose normal form will be henceforth denoted by $\tilde{\gr\nu}$), as
\be
\tilde\nu_j\ge 1 \; . \label{eigsepa}
\ee
The previous findings allow to recast such separability criteria
for $(m+n)$-mode states in terms of partially transposed symplectic invariants 
$\{\tilde{\Delta}_{j}\}$, defined by $\tilde{\Delta}_{j}\equiv 
M_{2j}(\Omega{\tilde{\gr\sigma}})$.
This simple result will be precious: 

\noindent{\em Little lemma.}
{Let ${\gr\sigma}$ be the physical CM of a state of a $(m+n)$-mode CV system, 
with $m\le n$. 
Let $\tilde{{\gr\sigma}}=T{\gr\sigma} T$ 
be the partial transposition of ${\gr\sigma}$ with respect to any of the two subsystems. 
Then, at most $m$ of the symplectic eigenvalues $\{\tilde{\nu}_j\}$ 
of $\tilde{\gr\sigma}$ can violate
Inequality (\ref{eigsepa}).}\smallskip

\noindent{\em Proof.} Suppose the transposition is performed in the $m$-mode subsystem.
Let ${\cal D}(\alpha)$ be the dimension of the subspace upon which 
the generic matrix $\alpha$ is negative definite. 
Since $T$ reduces to the identity on a $(2n+m)$-dimensional subspace, Inequality 
$\tilde{\gr\sigma}+i\Omega\ge0$
reduces to the (definitely satisfied) Inequality (\ref{obsheis}) on such a subspace, thus implying 
${\cal D}(\tilde{{\gr\sigma}}+i\Omega)\le m$. 
One has then ${\cal D}(\tilde{\gr\nu}+i\Omega)={\cal D}(\tilde{{\gr\sigma}}+i\Omega)\le m$,
because  
$\tilde{\gr\nu}+i\Omega=S^{T}(\tilde{{\gr\sigma}}+i\Omega)S$ 
for $S\in Sp_{2(m+n),\R}$ and the signature is preserved under congruence. 
The eigenvalues of 
$\tilde{\gr\nu}+i\Omega$ are given by $\{\tilde{\nu}_j\mp 1\}$, thus proving the lemma
as the $\{\tilde\nu_j\}$ have to be positive ($T{\gr\sigma} T>0$ because
${\gr\sigma}>0$). 
The choice of the transposed subsystem 
is not relevant, since the action by congruence of the matrix $\oplus_{1}^{m+n}s_z$ 
turns $\tilde{\gr\sigma}$ into the partial transpose under the $n$-mode subsystem 
$\tilde{\gr\sigma}'$ and 
$\Omega$ into $-\Omega$, and $\tilde{\gr\sigma}'-i\Omega\ge 0 \Leftrightarrow 
\tilde{\gr\sigma}'+i\Omega\ge 0$. $\Box$\smallskip

In analogy with \eq{sig}, let us define the transposed invariant  
$
\tilde{{\Sigma}}_{m+n} = \sum_{j=0}^{m+n} (-1)^{m+n+j} \tilde\Delta_j 
$.
The inequality 
\be
\tilde{{\Sigma}}_{m+n} \ge 0 \; , \label{symsepa}
\ee
being necessary for Inequality (\ref{eigsepa}) to be satisfied, 
{\em is a necessary condition for separability under $(m+n)$-mode bipartitions  
and is thus a sufficient condition to detect 
entanglement in such a multimode system, irrespective of the nature of the state under examination} 
\cite{nongausnote}.
Inequality (\ref{eigsepa}) cannot detect the negativity 
of the partial transpose 
whenever an even number of symplectic eigenvalues violates condition (\ref{eigsepa}).
However, because of the previous lemma, for $(1+n)$-mode Gaussian states  
(for which the PPT criterion is necessary and sufficient for separability \cite{werewolf}), 
at most one partially transposed symplectic eigenvalue can violate Inequality (\ref{eigsepa}). 
Inequality (\ref{symsepa}) is then always capable of detecting such a violation.
The same argument applies to `bisymmetric' Gaussian states, 
defined as the $(m+n)$-mode Gaussian states which are invariant under 
mode permutations internal to the $m$-mode and $n$-mode subsystems. 
A bisymmetric Gaussian state with CM $\gr\sigma$ can be reduced, by local symplectic operations 
(on the $m$-mode and $n$-mode subsytems), 
to the tensor product of a two-mode Gaussian state and of 
uncorrelated thermal states \cite{serafozzi05}, with global CM $\gr\sigma_{2}$:
$\gr\sigma_2 = S^{\sf T} \gr\sigma S$ for some $S\in\,Sp_{2m,\R}\oplus Sp_{2n,\R}$. 
The lemma above implies ${\cal D}(T\gr\sigma T+i\Omega)\le1$
from which, observing that
$
(TS^{\sf T}T) (T\gr\sigma T+i\Omega)(TST)  = 
T\gr\sigma_2 T+i\Omega
$, 
one infers that at most one 
partially transposed symplectic eigenvalue of the CM $\gr\sigma$ 
can violate Inequality (\ref{eigsepa}).
Notice that the locality of the operation $S$ is crucial in establishing this result, as it implies 
$(TS^{\sf T}T)\Omega (TST)=\Omega$. 

Summing up, {\em Inequality (\ref{symsepa}) is necessary and sufficient for the separability 
of $(1+n)$-mode and of bisymmetric $(m+n)$-mode Gaussian states} \cite{sternote}.
It has to be noted that
an effective strategy -- based on an iterative map -- 
to decide the separability of such states does exist \cite{giedkeprl01}. 
However, the condition (\ref{symsepa}) on the second moments 
can be readily analytically verified and 
may thus be very helpful in the detection of interesting CV entangled states.
This is especially relevant in view of the current experimental developments  
in the implementation of multipartite CV protocols \cite{yonezawa04},
which heavily rely on symmetric resources 
(perfectly discriminated by the previous condition).
A concrete example might better emphasize the usefulness of relation (\ref{symsepa}). 
The ``GHZ-type'' states considered in Refs.~\cite{vanloock00,vanloock03},
obtained by inserting squeezed vacua into an array of beam splitters, 
are the prototypical 
resource for the implementation of teleportation networks. 
Moreover, they turn out to be 
the symmetric Gaussian states maximising both the couplewise (between any pair of modes) 
and the genuine multipartite entanglement \cite{adesso05}.
To fix ideas, let us consider a four mode state [whose CM is given by Eq.~(31) 
of Ref.~\cite{vanloock03} for $N=4$] with squeezing parameters $r_1=r_2=r$ 
(see the notation of Ref.~\cite{vanloock03}). 
Realistically, thermal noise with mean photon number $q-1$ 
will also be assumed to affect the creation of such states
(this amounts to multiplying the CM by $q$). 
Here, two kinds of bipartitions are of interest: 
the one between two modes and the remaining two and the one between 
one mode and the other three.
Working out the partially transposed symplectic 
eigenvalues to check the separability of such bipartitions
would require the numerical solution of an eighth degree
algebraic equation. Instead, Inequality (\ref{symsepa}) allows to analytically verify the separability 
of any bipartition with the elementary operations (multiplication and addition) 
needed to compute the minors $\tilde\Delta_j$.
For instance, for $2\times 2$ bipartitions one gets $\tilde\Delta_j=q^{2j}(1+g\cosh(4r))(C_{j}^4-2g)$,
where $g=\min(j,4-j)$ and $C_j^4$ is the binomial coefficient.
Checking condition (\ref{symsepa}) is then straightforward and leads to the following 
analytical relation for the separability of the considered bipartiton: $\cosh(4r) \le (q^4+1)/(2q^2)$.
Clearly, the advantage of such a strategy 
gets more and more relevant as the number of modes increases.

The compact and elegant forms in which uncertainty relations 
and separability criteria (remarkably necessary and sufficient for two relevant
classes of Gaussian states) have been recast in the present paper
were obtained relying exclusively on symplectic analysis applied at the phase space level.
Much scope is left to such a 
longstanding \cite{simon8794} kind of approach, in
particular concerning the entanglement characterization of CV states. 
The analysis of Gaussian states based on symplectic invariants 
has provided remarkable insight into the entanglement properties of two-mode
states \cite{adesso03} and 
could be, employing the techniques developed in this paper, 
extended to multimode settings and to the analysis of multipartite continuous variable entanglement
\cite{vanloock03,yonezawa04,vanloock00,giedke01},
which has been lately drawing considerable attention 
\cite{wolf04,adesso04,adesso05,zhang05}.
At a more fundamental level,
this paper is intended to shed further light on the constraints imposed by quantum mechanics 
on second moments of canonical operators 
and on the symplectic structure 
underlying the evolution of bosonic modes under quadratic Hamiltonians.

\acknowledgments

I would like to thank: G.~Adesso, F.~Illuminati, 
S.~De Siena, J.~Eisert, S.~Bose, J. van Enk, T.~Coudreau, M.~Tondin, 
M.B.~Plenio, S.~Bergamini, J.L.~Cumming and J.E.J.~Morton.
This research is part of QIP IRC www.qipirc.org (GR/S82176/01).


\begin{thebibliography}{99}
\bibitem{ccrnote} The CCR could be expressed in terms of 
the (bounded) operators $D_{\xi}=\exp{(i\hat R^{\sf T}\Omega\xi)}$, 
for $\xi\in\R^{2n}$, as $D_\xi^{\dag}D_{\xi'}=D_{\xi'}D_{_xi}^{\dag}\exp{(4i\xi^{\sf T}\Omega\xi')}$.
However, to our aims, the simpler form $[\hat{R}_j,\hat{R}_k]=\Omega_{jk}$ will suffice.
Through all the paper, the second moments of the operators
$\{\hat R_j\}$ are undestood to be finite for the considered states.
\bibitem{minornote}The principal minor of order $k$ of the $n\times n$ matrix
$\alpha$, $M_k(\alpha)$, 
is the sum of the determinants of all the $k\times k$ 
submatrices of $\alpha$ obtained by deleting $n-k$ rows and the corresponding 
$n-k$ columns.
\bibitem{nongausnote}Let us remark that for non Gaussian states Inequality (\ref{eigsepa}) 
on second moments is necessary but not sufficient for PPT. 
For such states, conditions on higher moments allow for a sharper 
detection of entanglement \cite{agarwal05}.
\bibitem{sternote}This holds for all states, except for the set of null measure 
for which any $\tilde{\nu}_j$ is identical to $1$ 
(that could be entangled but have $\tilde\Sigma_{n} = 0$). 
However, even for such states 
a necessary and sufficient condition for separability may always be found in terms of the 
quantities $\tilde{\Delta}_{j}$. 
If $\tilde\nu_j=1$ for $p$ eigenvalues, this condition reads:
$\sum_{j=p}^{m+n} \left[(-1)^{m+n+j} p\,! C^j_p 
\tilde{\Delta}_{j} \right]  \ge 0$ (see \cite{mestesso} for a proof).
\end{thebibliography}
\end{document}